\documentclass[prb,a4paper,10pt,twocolumn,floatfix,superscriptaddress,amsmath,amsfonts,amssymb,preprintnumbers,citeautoscript]{revtex4}
\pdfoutput=1
\usepackage[T1]{fontenc}\usepackage[latin1]{inputenc}
\usepackage{dcolumn,graphicx,color,booktabs,microtype,afterpage} 
\usepackage{sidecap}
\usepackage{times}
\usepackage[colorlinks,plainpages=false,linkcolor=blue,urlcolor=blue,citecolor=blue,pdfpagemode=UseNone,pdfstartview=FitBH]{hyperref}

\usepackage{amssymb}
\usepackage{amsmath}
\usepackage{mathtools}
\usepackage{graphicx}

\newcommand{\ersio}{C-Er$_{2}$Si$_{2}$O$_{7}$}
\newcommand{\dersio}{D-Er$_{2}$Si$_{2}$O$_{7}$}
\newcommand{\ybsio}{Yb$_{2}$Si$_{2}$O$_{7}$}

\newcommand{\invang}{\AA$^{-1}$}

\begin{document}
%
%
%
%
\title
{N\'eel Ordering in the Distorted Honeycomb Pyrosilicate: C-Er$_{2}$Si$_{2}$O$_{7}$}
\author{Gavin Hester}
\email{Gavin.Hester@colostate.edu}
\affiliation{Department of Physics, Colorado State University, 200 W. Lake St., Fort Collins, CO 80523-1875, USA}
\author{T. N. DeLazzer}
\affiliation{Department of Physics, Colorado State University, 200 W. Lake St., Fort Collins, CO 80523-1875, USA}
\author{S. S. Lim}
\affiliation{Department of Physics, Colorado State University, 200 W. Lake St., Fort Collins, CO 80523-1875, USA}
\author{C. M. Brown}
\affiliation{NIST Center for Neutron Research, National Institute of Standards and Technology, Gaithersburg, Maryland, 20899-6102, USA}
\author{K. A. Ross}
\email{Kate.Ross@colostate.edu}
\affiliation{Department of Physics, Colorado State University, 200 W. Lake St., Fort Collins, CO 80523-1875, USA}
\affiliation{Quantum Materials Program, CIFAR, Toronto, Ontario M5G 1Z8, Canada}
\date{\today}
%
%
%
%
\begin{abstract}
The rare-earth pyrosilicate family of compounds (RE$_{2}$Si$_{2}$O$_{7}$) hosts a variety of polymorphs, some with honeycomb-like geometries of the rare-earth sublattices, and the magnetism has yet to be deeply explored in many of the cases. Here we report on the ground state properties of \ersio. \ersio\ crystallizes in the C2/m space group and the Er$^{3+}$ atoms form a distorted honeycomb lattice in the $a$-$b$ plane. We have utilized specific heat, DC susceptibility, and neutron diffraction measurements to characterize \ersio. Our specific heat and DC susceptibility measurements show signatures of antiferromagnetic ordering at 2.3 K. Neutron powder diffraction confirms this transition temperature and the relative intensities of the magnetic Bragg peaks are consistent with a collinear N\'eel state in the magnetic space group C2'/m, with ordered moment of 6.61 $\mu_{B}$ canted 13$^{\circ}$ away from the $c$-axis toward the $a$-axis. These results are discussed in relation to the isostructural quantum dimer magnet compound \ybsio.
\end{abstract}

\maketitle

%
%
%
%
%

\section{Introduction}
Recent efforts in the study of quantum magnetism and novel magnetic ground states have focused on the use of 4\textit{f} magnetic atoms in different frustrated geometries (i.e. triangular \cite{Sarkar2019, yahne2020pseudospin, Paddison2016}, kagome \cite{ashtar2020new}, and pyrochlore lattices \cite{rau2019frustrated}) instead of the traditional 3\textit{d} magnetic atoms, such as Cu$^{2+}$ or Ni$^{2+}$. This is due to numerous advantages to 4\textit{f} magnetic atoms over traditional 3\textit{d} magnetic atoms. One of these advantages is 4\textit{f} ions can often be interchanged for each other within the same structure, which tends to produce a wide variety magnetic behaviors. Additionally - at the single-ion level - the effect of strong spin-orbit coupling of the 4\textit{f} ions and the surrounding crystal electric field (CEF) is often to produce an energetically well-isolated doublet that can be mapped to an effective spin $\frac{1}{2}$ ($S_{\text{eff}}=1/2$). These $S_{\text{eff}}=1/2$ systems have been shown to exhibit many of the same quantum ground states expected for ``bare'' spin $\frac{1}{2}$ systems \cite{Rau2018, Wu2019, Hester2019}. The highly localized 4\textit{f} electrons are also advantageous as they lead to weak orbital overlap, and therefore weak superexchange (typically on the order of 1~K). This weak superexchange drives magnetic ordering transitions down in temperature, but also allows for field-induced transitions to generally be accessible with conventional, widely-available superconducting magnets. Finally, the orbitally-active effective spins allow for bond-dependent exchange interactions, which can yield novel quantum phases such as the Kitaev quantum spin liquid \cite{Kitaev2006, Jackeli2009} or Quantum Spin Ice \cite{Ross2011, hermele2004pyrochlore}. The Kitaev model was originally derived for the honeycomb lattice and is therefore of particular relevance to this work, as \ersio\ forms a distorted honeycomb lattice of Er$^{3+}$ ions in the $a$-$b$ plane. Numerous magnetic honeycomb systems have been investigated in the context of the Kitaev model \cite{Banerjee2016, choi2019exotic, Jackeli2013, HwanChun2015}, but the quantum spin liquid state of the pure Kitaev model has yet to be discovered in a real material. This provides impetus to study new honeycomb materials with strong spin-orbit coupling, which will aid theoretical and materials design efforts towards a Kitaev quantum spin liquid.

\begin{figure}[htp]
\includegraphics[scale = 0.8]{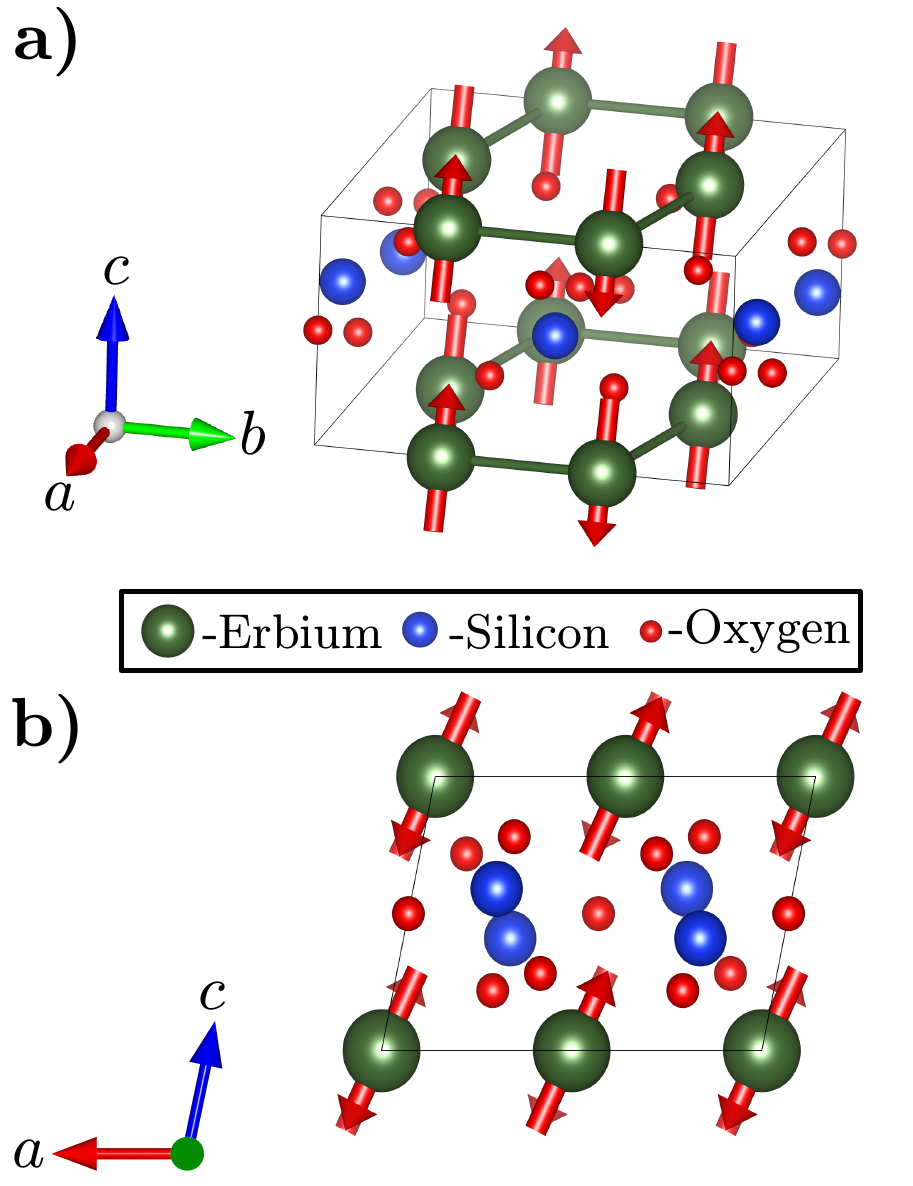}
\caption{a) The crystal and magnetic structure (obtained from the refinement in Fig.~\ref{fig_diffraction}a) of \ersio. The Er$^{3+}$ atoms form a distorted honeycomb lattice in the $a$-$b$ plane. Here Er atoms are green, Si are blue, and O are red. b) View of the crystal and magnetic structure along the $b$-axis showing the separation between the layers. The interlayer distance is 4.722 \AA, which is slightly larger then the nearest-neighbor bond distance of 3.477 \AA. All panels of this figure were created using the Vesta software \cite{vesta}.} \label{fig_crystals}
\end{figure}

The rare-earth pyrosilicate family (RE$_{2}$Si$_{2}$O$_{7}$) of compounds was synthesized in the 1980's by Felsche \cite{Felsche1970}. Structures were determined for all of the lanthanide atoms in this series, with many of the lanthanides exhibiting a diamorphic or polymorphic phase diagram. The interest in these compounds was primarily due to the Si-O-Si bonds in the [Si$_{2}$O$_{7}$]$^{6-}$ groups. Therefore, little work was done on the low temperature magnetic properties of the series at the time. The Er-based pyrosilicate compound Er$_{2}$Si$_{2}$O$_{7}$ can crystallize in three different structures depending on the synthesis temperature: the low-temperature phase P1 (Type B), the intermediate-temperature phase C2/m (Type C) and the high-temperature phase P2$_{1}$/a (Type D) \cite{Felsche1970, smolin1970crystal}. We have recently begun exploring the low-temperature properties of some members of the rare-earth pyrosilicate series. In particular, we have studied \dersio, which shows evidence of being a new experimental platform for studying the transverse-field Ising model \cite{Hester2020}. Of more relevance to the current work, we have also studied the magnetic properties of \ybsio \cite{Hester2019}, which forms the same C-type "thortveitite" structure as the title compound, \ersio. We also note that Lu$_{2}$Si$_{2}$O$_{7}$, Y$_{2}$Si$_{2}$O$_{7}$ \cite{redhammer2003beta} and Tm$_{2}$Si$_{2}$O$_{7}$ \cite{kahlenberg2014thortveitite} form the same structure.  Lu$_{2}$Si$_{2}$O$_{7}$ and Y$_{2}$Si$_{2}$O$_{7}$ are non-magnetic, so they may provide a useful non-magnetic analog to C-Er$_{2}$Si$_{2}$O$_{7}$.  Meanwhile, the magnetism of Tm$_{2}$Si$_{2}$O$_{7}$ has not been thoroughly investigated, though it does show a low-temperature Schottky-like anomaly in the specific heat \cite{ciomaga2020optical}, similar to \ybsio. \ybsio\ does not magnetically order down to 50 mK in zero field, and exhibits a field-induced phase similar to the triplon Bose-Einstein condensates observed in 3\textit{d} transition metal based dimer magnets. This brings us to the compound of current interest, \ersio. \ersio\ is isostructural to \ybsio, thus providing an opportunity to study how rare-earth substitution influences the magnetic ground state properties of quantum magnets with the same structures. In this thortveitite structure, the Er$^{3+}$ ion resides in a distorted octahedral site at the 4$g$ Wyckoff position, resulting in a CEF with C$_{2}$ point group symmetry. The Er-O bond lengths in this structure range from 2.234 - 2.279 \AA. The [ErO$_{6}$] groups share three of their edges with adjacent groups, forming the honeycomb layers in the $a$-$b$ plane. The magnetic transition temperature of \ersio\ has been reported to be 2.50(5) K previously \cite{Maqsood1979,Maqsood1981} - although the data was not shown - and magnetization measurements performed above room temperature have been reported \cite{Ameer2012}. No further studies of the magnetic properties have been undertaken until this work.

\section{Experimental Methods}
Powder samples of \ersio\ were synthesized by grinding stoichiometric amounts of SiO$_{2}$ and Er$_{2}$O$_{3}$, pressing the reactants into a rod, and heating 8 times at 1300$^{\circ}$C for 48 hours with thorough grinding between heatings. Phase purity was confirmed using powder x-ray diffraction. Refinement of 4 K neutron diffraction data yielded the lattice parameters: $a$ = 6.8529(4) \AA, $b$ = 8.9446(5) \AA, $c$ = 4.7219(3) \AA, $\alpha$ = 90$^{\circ}$, $\beta$ = 101.763(4)$^{\circ}$, $\gamma$ = 90$^{\circ}$. These parameters are consistent with previously published values \cite{Felsche1970}. \textcolor{red}{} The crystal structure and magnetic structure of \ersio\ are shown in Fig.~\ref{fig_crystals}.

Specific heat measurements were performed on a 1.1 mg piece of sintered powder \ersio\ using a Quantum Design PPMS \footnote{Certain commercial equipment, instruments, or materials are identified in this document. Such identification does not imply recommendation or endorsement by the National Institute of Standards and Technology, nor does it imply that the products identified are necessarily the best available for the purpose.} with a dilution insert. Temperature dependent susceptibility measurements were performed on a powder sample of \ersio\ from the same batch, immobilized in eicosane wax, in a Quantum Design MPMS-XL system with a 100 Oe DC magnetic field. Field dependent magnetization measurements were performed on the same sample at a temperature of 1.8 K. Powder neutron diffraction measurements were performed on the BT-1 high resolution powder diffractometer at the NIST Center for Neutron Research with 60' collimation and the Ge(311) monochromator ($\lambda = 2.079$ \AA). The neutron diffraction sample consisted of 3.5 grams of \ersio\ powder loaded in an aluminum sample canister, with 1 bar of helium exchange gas loaded at room temperature. The sample was cooled using a $\prescript{3}{}{\textnormal{He}}$ refrigerator, and measurements were performed between 0.3 and 4~K. All error bars shown in this work indicate one standard deviation.

\begin{figure}[!t]
\includegraphics[scale = 0.6]{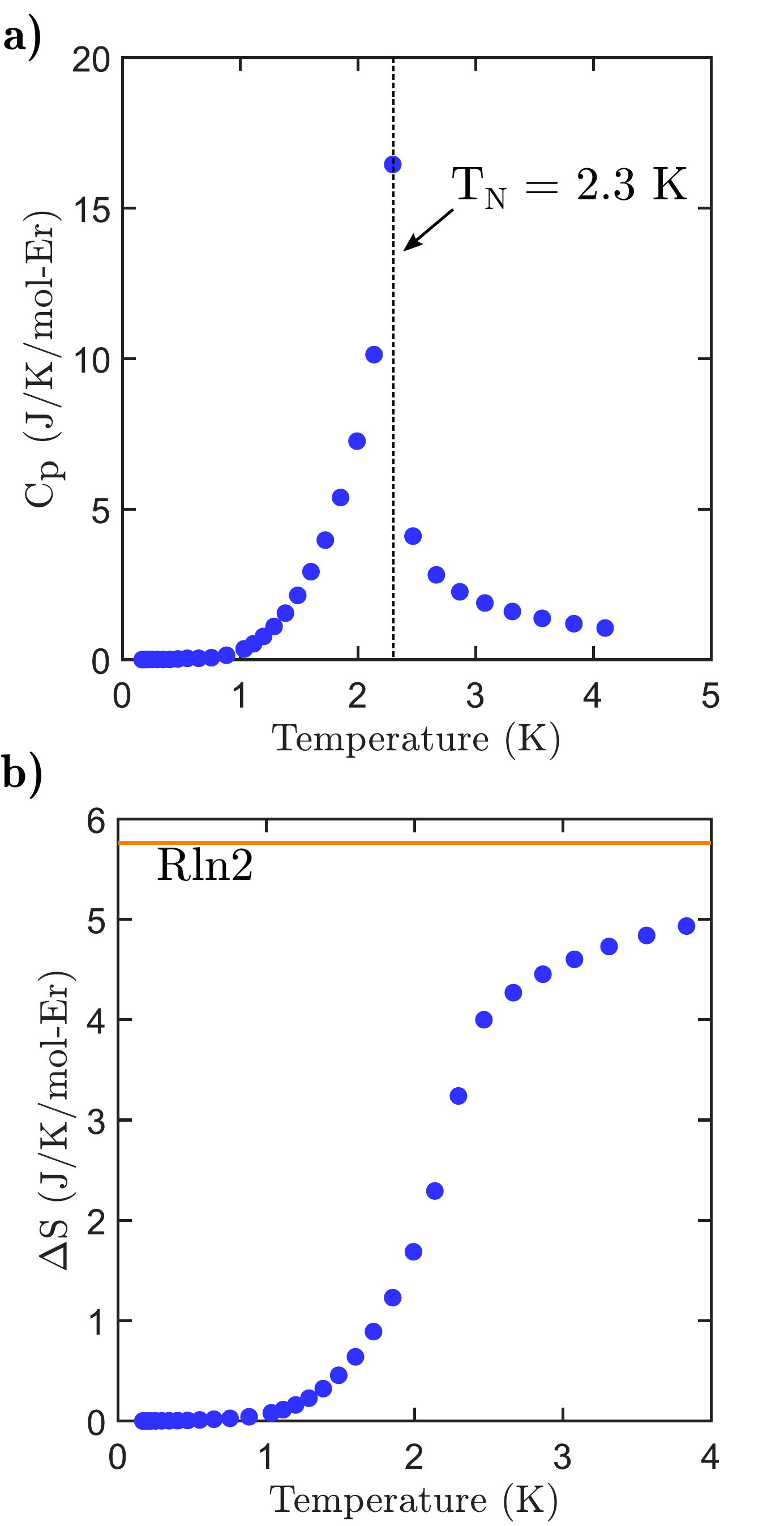}
\caption{a) Heat capacity vs. temperature, $C_p(T)$, of a powder sample of \ersio\ using a Quantum Design PPMS with dilution insert. A lambda anomaly is observed at T$_{N}$ = 2.3 K, indicative of a transition to long range magnetic order. b) Entropy calculated from the C$_{p}$ vs. T shown in panel a. The entropy approaches $R$ln2 through the transition, consistent with what one would expect for an S$_{\text{eff}}$ = 1/2 system.} \label{fig_cp}
\end{figure}

\begin{figure*}[htp]
\includegraphics[scale = 0.42]{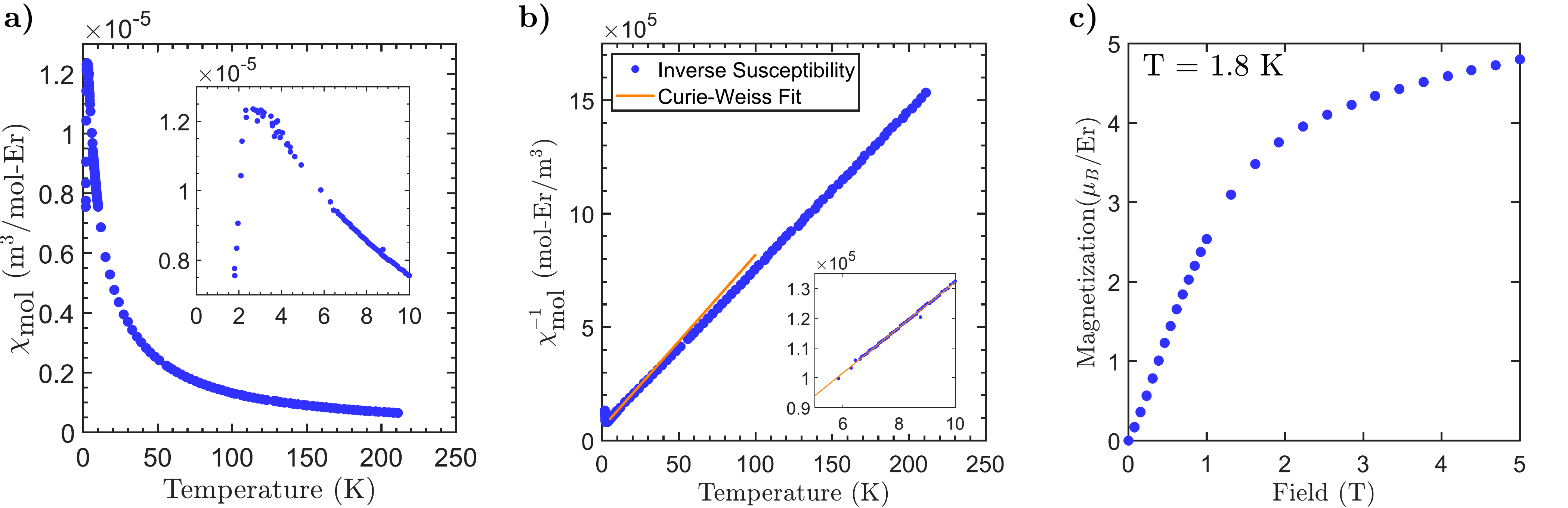}
\caption{a) Susceptibility vs. temperature, $\chi(T)$, of a powder sample of \ersio\ in an applied field of 0.01 T. A sharp drop-off of the susceptibility occurs at 2.3 K, consistent with antiferromagnetic order. Susceptibility values are shown in SI units (m$^{3}$/mol-Er), the value in Gaussian units (emu/mol-Er) can be obtained by multiplying by 4$\pi$ x 10$^{-6}$. (inset) Zoom-in of the low temperature region of $\chi(T)$. b) $\chi^{-1}(T)$ and fit of the Curie-Weiss law. The fit was limited to the temperature range of 5 - 10 K, yielding an effective moment of $\mu_{\text{eff}} = $ 9.1(5) $\mu_{B}$, and a Curie-Weiss temperature of $\theta_{CW} = -7.3(2)$ K. (inset) Zoom in on the region where the Curie-Weiss fit was performed. c) Field dependent magnetization measurements performed at 1.8 K on a powder of \ersio. At 5 T the magnetization appears to be approaching the saturated limit, and powder-averaged moment of 4.8 $\mu_{B}$ is observed.
} \label{fig_magnetometry}
\end{figure*}

\section{Results and Discussion}
\subsection{Specific Heat and Magnetization}
Specific heat data obtained between 150 mK and 4~K (Fig.~\ref{fig_cp}a) show a lambda anomaly at $T_N = 2.3$ K, indicating a transition to long range magnetic order. This transition temperature is close to the previously reported 2.50(5) K \cite{Maqsood1979,Maqsood1981}, though the previous reports do not mention the type of measurement used to determine this or show any data. The entropy calculated from the specific heat is shown in Fig.~\ref{fig_cp}b. The entropy approaches $R$ln2 - the expected value for an $S_{\text{eff}}$ = 1/2 system - but falls short of it, likely indicating that some short range correlations persist to temperatures higher than 4 K. The transition temperature is corroborated by the observation of a peak in the susceptibility, shown in Fig.~\ref{fig_magnetometry}a-b. The sharp downturn in the susceptibility after the transition indicates the system enters an antiferromagnetic ground state. A Curie-Weiss fit to the inverse susceptibility is shown in Fig.~\ref{fig_magnetometry}b. The fit was performed only for temperatures between 5 and 10 K, avoiding lower temperatures due to the onset of significant correlations, and higher temperatures due to the (likely) mixing of higher CEF levels. This fit yielded a Curie constant of 10.4(1) mol-Er$\cdot$K/emu (8.30(12) $\times$ 10$^{5}$ mol-Er$\cdot$K/m$^{3}$), corresponding to an effective moment of $\mu_{\text{eff}}$ =9.1(5) $\mu_{B}$, which is below the free-ion value $\mu_{\text{eff}}^{\text{free}} = g_J \sqrt{J(J+1)} = 9.58 \mu_{B}$, where $J=15/2$ is the total angular momentum and $g_J=6/5$ is the Lande $g$-factor for Er$^{3+}$. The Curie-Weiss temperature was $\theta_{CW}$=-7.3(2) K. This yields a frustration index ($f = \theta_{CW}/T_{N}$) of 3.2, indicating the system is only lightly frustrated. However, we note that Curie-Weiss fits can prove unreliable for rare-earth ions, with significant non-linearity in the inverse susceptibility arising from CEF effects. As the first crystal field level energy is not known for \ersio, it is uncertain how accurate this fit is, even within this low temperature regime. 

The field dependence of the magnetization at $T$=1.8 K is shown in Fig.~\ref{fig_magnetometry}c. The magnetization is approximately linear up to 1 T, where it then begins to saturate near 5 T, though we note that a fully saturated state is not expected due to mixing with higher CEF levels as the field is increased. The powder averaged ``saturated'' moment at 5~T is nearly 4.8 $\mu_{B}$, which falls well below the maximum allowed value of $\mu_{\text{eff}}^{\text{free}} = 9.58 \mu_{B}$. This is not surprising, as the Kramer's doublet CEF ground state likely carries a strongly anisotropic $g$-factor, which will appear as a reduced saturated moment on powder averaged data. Further, the ground state CEF wavefunctions likely do not solely consist of the maximum $J_z$ eigenstates, which will also reduce the observed magnetic moment compared to the maximum possible value.

\begin{figure*}[htp]
\includegraphics[scale = 0.9]{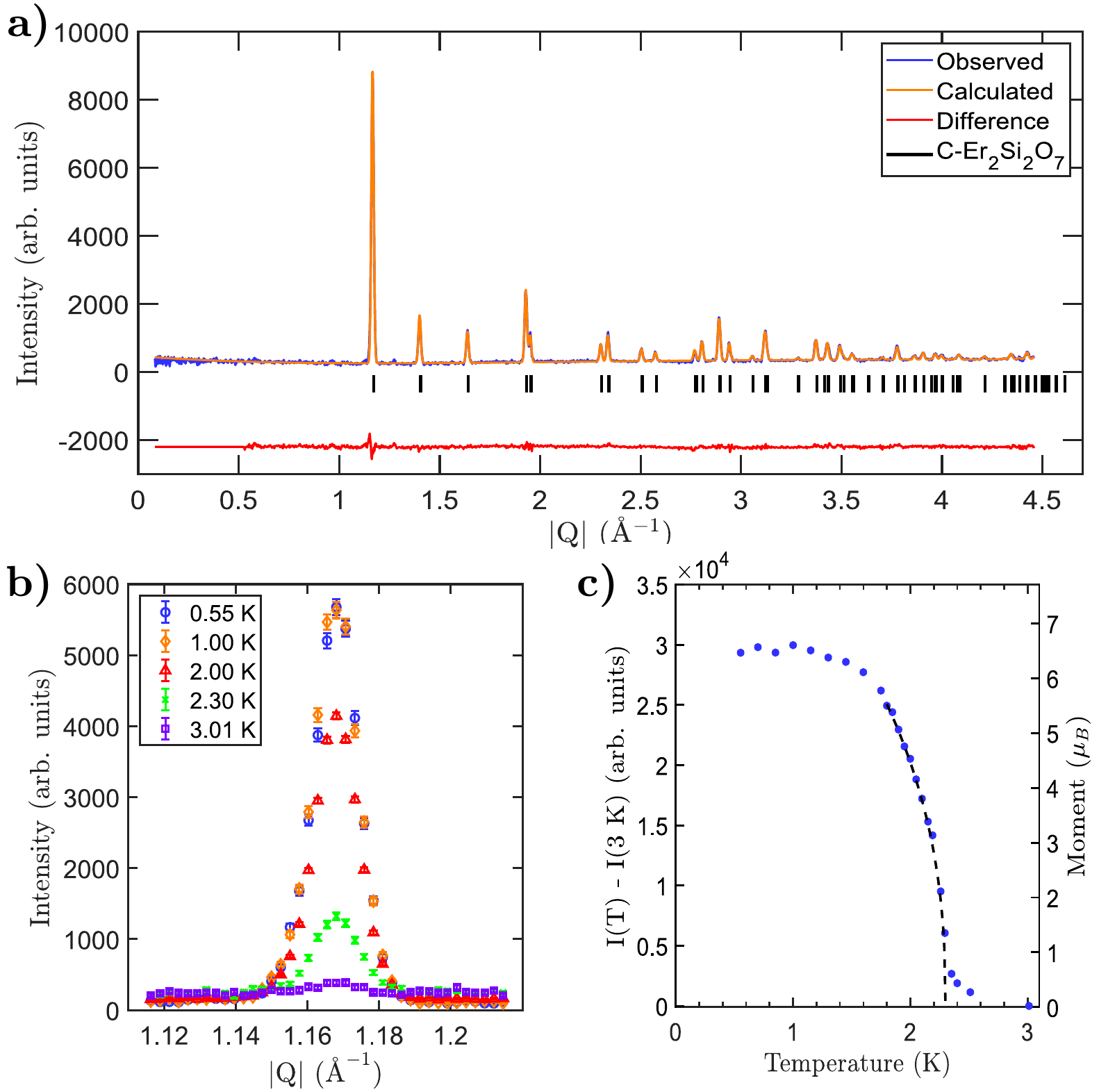}
\caption{a) Rietveld refinement of neutron diffraction data measured at 320 mK with the 4 K data (nuclear Bragg peaks) subtracted, which is well fit by the $|$$\vec{k}$$|$ = 0 structure shown in Fig.~\ref{fig_crystals}. b) A selection of peaks from parametric scans over the (110) Bragg peak, used for the order-parameter curve in panel c. c) Evolution of the (110) Bragg peak intensity with temperature. The black dashed line is a guide to the eye showing the extrapolation to the N\'eel temperature of 2.3 K, as expected based on the thermodynamic measurements above. The right-side axis shows the corresponding ordered moment, determined based on the refinement of the 320 mK data. Error bars are smaller than the points shown. Note: each intensity point shown is the sum of the intensity over the whole peak between Q = [1.11, 1.21] \invang.} \label{fig_diffraction}
\end{figure*}

\subsection{Neutron Diffraction}
Powder neutron diffraction measurements obtained using BT-1 were refined using the FullProf software \cite{fullprof} and the SARAh suite (using the Kovalev tables) \cite{sarah1, sarah2}. The resulting Rietveld refinement for the magnetic structure of \ersio\ on the data obtained at 320 mK with 4 K data subtracted is shown in Fig.~\ref{fig_diffraction}a. A symmetry analysis of the allowed $|$$\vec{k}$$|$ = 0 magnetic structures for \ersio\ provided four possible irreducible representations. The Rietveld refinement of \ersio\ yielded a $|$$\vec{k}$$|$ = 0 magnetic structure in the $\Gamma_{4}$ irreducible representation with the space group C2$^\prime$/m. The $\Gamma_{4}$ irreducible representation contained two basis vectors for atom 1: $\psi_{5}$ = (2,0,0) and $\psi_{6}$ = (0, 0, 2). The basis vectors for atom 2 are the negative of the basis vectors for atom 1, guaranteeing colinear antiferromagnetism. The coefficients for these basis vectors are $C_{5}$ = -0.84(3) and $C_{6}$ = 3.03(1). These coefficients yield the magnetic structure shown in Fig.~\ref{fig_crystals}. The structure is a simple N\'eel structure with the moments canted 13$^{\circ}$ away from the $c$-axis toward the $a$-axis. Multiplying the coefficients by their respective basis vectors and taking into account the non-orthogonal axes, the ordered moment extracted from the fit is 6.61 $\mu_{B}$.

The thermal evolution of the (110) magnetic Bragg peak intensity is shown in Fig.~\ref{fig_diffraction}b-c. The data shown in Fig.~\ref{fig_diffraction}c is a sum of the intensity over the full detector sweep used (i.e. the range of $|Q|$ values shown for the example peaks in panel b). The magnetic intensity shows the typical shape for a second order antiferromagnetic transition with a N\'eel temperature of 2.3 K. An attempt was made to fit the intensity in the proximity of the N\'eel temperature to extract the order parameter's critical exponent, $\beta$. However, the point density in the proximity of T$_{N}$ was too low to accurately determine the critical exponent. Additionally, there is critical scattering that persists above T$_{N}$ that further complicates the fitting of the critical exponent. The dashed line on this figure is a guide to the eye.

\subsection{Comparison to Yb$_{2}$Si$_{2}$O$_{7}$}
As \ersio\ is isostructural to \ybsio\ , it is worthwhile to consider why \ersio\ magnetically orders and \ybsio\ does not (in the absence of a magnetic field). The difference in ground states may be due to slight differences in bond lengths. The nearest-neighbor and next-nearest-neighbor bond lengths differ by 3.5\% in \ybsio\ , while the difference is only 2.5\% for \ersio. This indicates that \ersio\ is closer, at least in the sense of the bond lengths, to being a ``perfect'' honeycomb lattice. It is well-known that the ground-state of a structurally isotropic honeycomb lattice with nearest-neighbor antiferromagnetic XXZ exchange is the N\'eel state \cite{Fouet2001}, due to the bipartite nature of the lattice. Likewise, the presence of more pronounced structural dimers in \ybsio\ compared to \ersio\ would naturally seem to lead to a dimerized quantum (singlet) state. However, the details of the exchange interactions in rare earth insulators is not only a matter of bond lengths - it also depends on the bond angles and the nature of the angular momentum states involved in the exchange \cite{Rau2018Yb}. We therefore expect some differences between \ersio\ and \ybsio\ to arise due to the presence of different strengths of anisotropic exchange. It would be reasonable to expect dipolar interactions to be stronger in \ersio\ due to the higher overall moment Er$^{3+}$ ($\mu_{\text{eff}} = 9.1$) in comparison to that of Yb$^{3+}$ ($\mu_{\text{eff}} = 3.0 \mu_{B}$) in this structure. Ideally, the exchange interactions of \ersio\ could be extracted via high-field inelastic neutron scattering, but this requires single crystal samples. We have not been successful in growing single crystals of \ersio\ via the optical floating zone technique because the high-temperature polymorph, \dersio, is stabilized from the melt instead \cite{Nair2019}.

\section{Conclusions}
We have used specific heat, magnetic susceptibility, and neutron diffraction measurements to explore the low-temperature properties and magnetic structure of \ersio. Our specific heat measurements show a sharp lambda-like anomaly at T$_{N}$ = 2.3 K indicating a transition to long range magnetic order. This is corroborated by magnetic susceptibility measurements and neutron powder diffraction, which both reveal an antiferromagnetic transition at 2.3 K. Furthermore, the neutron diffraction measurements allowed for the determination of the magnetic structure below 2.3 K. The ordered magnetic structure consists of a $|k|=0$ antiferromagnetic N\'eel state, with 6.8 $\mu_B$ moments pointing in the $a$-$c$ plane (13$^{\circ}$ from $c$). The different ground states of \ersio\ and the isostructural quantum dimer magnet system \ybsio\ are interesting from the perspective of understanding how the relevant rare-earth species affects the nature of exchange interactions in quantum magnets of fixed geometry. One possible avenue for future work, which would not require single crystals, is the measurement of the crystal field parameters for Er$^{3+}$ in \ersio. The C$_{2}$ point group symmetry results in eight independent Steven's parameters. These can likely be determined via inelastic neutron scattering, since there are also eight Kramer's doublets for the Er$^{3+}$ ion ($J=\frac{15}{2}$), thus their energies and relative intensities provide enough constraints to determine the crystal field Hamiltonian. This would also allow some insight into the crystal field Hamiltonian for \ybsio, an important open question that cannot be easily addressed using inelastic neutron scattering directly due to there being fewer Kramer's doublets available (only four in that case), which can result in an under-constrained fit \cite{sarkis2020}. An additional avenue for future work would be the determination of the exchange interactions for \ersio\ using inelastic neutron scattering on single crystal samples, if they become available. This would be of great benefit to further enable the understanding of how different rare-earth species affect the nature of the exchange interactions in quantum magnets.

\section{Acknowledgements}
This research was supported by the National Science Foundation Agreements No. DMR-1611217 and DMR-2005143. G. Hester would like to acknowledge D. R. Yahne and C. L. Sarkis for helpful discussions and support. The authors acknowledge the Central Instrument Facility at Colorado State University for instrument access and training. We acknowledge the support of the National Institute of Standards and Technology, U. S. Department of Commerce, in providing the neutron research facilities used in this work.

\bibliographystyle{apsrev}

\begin{thebibliography}{33}
\expandafter\ifx\csname natexlab\endcsname\relax\def\natexlab#1{#1}\fi
\expandafter\ifx\csname bibnamefont\endcsname\relax
  \def\bibnamefont#1{#1}\fi
\expandafter\ifx\csname bibfnamefont\endcsname\relax
  \def\bibfnamefont#1{#1}\fi
\expandafter\ifx\csname citenamefont\endcsname\relax
  \def\citenamefont#1{#1}\fi
\expandafter\ifx\csname url\endcsname\relax
  \def\url#1{\texttt{#1}}\fi
\expandafter\ifx\csname urlprefix\endcsname\relax\def\urlprefix{URL }\fi
\providecommand{\bibinfo}[2]{#2}
\providecommand{\eprint}[2][]{\url{#2}}

\bibitem[{\citenamefont{Sarkar et~al.}(2019)\citenamefont{Sarkar, Schlender,
  Grinenko, Haeussler, Baker, Doert, and Klauss}}]{Sarkar2019}
\bibinfo{author}{\bibfnamefont{R.}~\bibnamefont{Sarkar}},
  \bibinfo{author}{\bibfnamefont{P.}~\bibnamefont{Schlender}},
  \bibinfo{author}{\bibfnamefont{V.}~\bibnamefont{Grinenko}},
  \bibinfo{author}{\bibfnamefont{E.}~\bibnamefont{Haeussler}},
  \bibinfo{author}{\bibfnamefont{P.~J.} \bibnamefont{Baker}},
  \bibinfo{author}{\bibfnamefont{T.}~\bibnamefont{Doert}}, \bibnamefont{and}
  \bibinfo{author}{\bibfnamefont{H.-H.} \bibnamefont{Klauss}},
  \bibinfo{journal}{Phys. Rev. B} \textbf{\bibinfo{volume}{100}},
  \bibinfo{pages}{241116} (\bibinfo{year}{2019}).

\bibitem[{\citenamefont{Yahne et~al.}(2020)\citenamefont{Yahne, Sanjeewa,
  Sefat, Stadelman, Kolis, Calder, and Ross}}]{yahne2020pseudospin}
\bibinfo{author}{\bibfnamefont{D.~R.} \bibnamefont{Yahne}},
  \bibinfo{author}{\bibfnamefont{L.~D.} \bibnamefont{Sanjeewa}},
  \bibinfo{author}{\bibfnamefont{A.~S.} \bibnamefont{Sefat}},
  \bibinfo{author}{\bibfnamefont{B.~S.} \bibnamefont{Stadelman}},
  \bibinfo{author}{\bibfnamefont{J.~W.} \bibnamefont{Kolis}},
  \bibinfo{author}{\bibfnamefont{S.}~\bibnamefont{Calder}}, \bibnamefont{and}
  \bibinfo{author}{\bibfnamefont{K.~A.} \bibnamefont{Ross}},
  \bibinfo{journal}{Phys. Rev. B} \textbf{\bibinfo{volume}{102}},
  \bibinfo{pages}{104423} (\bibinfo{year}{2020}).

\bibitem[{\citenamefont{Paddison et~al.}(2016)\citenamefont{Paddison, Daum,
  Dun, Ehlers, Liu, Stone, Zhou, and Mourigal}}]{Paddison2016}
\bibinfo{author}{\bibfnamefont{J.~A.~M.} \bibnamefont{Paddison}},
  \bibinfo{author}{\bibfnamefont{M.}~\bibnamefont{Daum}},
  \bibinfo{author}{\bibfnamefont{Z.}~\bibnamefont{Dun}},
  \bibinfo{author}{\bibfnamefont{G.}~\bibnamefont{Ehlers}},
  \bibinfo{author}{\bibfnamefont{Y.}~\bibnamefont{Liu}},
  \bibinfo{author}{\bibfnamefont{M.~B.} \bibnamefont{Stone}},
  \bibinfo{author}{\bibfnamefont{H.}~\bibnamefont{Zhou}}, \bibnamefont{and}
  \bibinfo{author}{\bibfnamefont{M.}~\bibnamefont{Mourigal}},
  \bibinfo{journal}{Nature Physics} \textbf{\bibinfo{volume}{13}},
  \bibinfo{pages}{117} (\bibinfo{year}{2016}).

\bibitem[{\citenamefont{Ashtar et~al.}(2020)\citenamefont{Ashtar, Guo, Wan,
  Wang, Gong, Liu, Su, and Tian}}]{ashtar2020new}
\bibinfo{author}{\bibfnamefont{M.}~\bibnamefont{Ashtar}},
  \bibinfo{author}{\bibfnamefont{J.}~\bibnamefont{Guo}},
  \bibinfo{author}{\bibfnamefont{Z.}~\bibnamefont{Wan}},
  \bibinfo{author}{\bibfnamefont{Y.}~\bibnamefont{Wang}},
  \bibinfo{author}{\bibfnamefont{G.}~\bibnamefont{Gong}},
  \bibinfo{author}{\bibfnamefont{Y.}~\bibnamefont{Liu}},
  \bibinfo{author}{\bibfnamefont{Y.}~\bibnamefont{Su}}, \bibnamefont{and}
  \bibinfo{author}{\bibfnamefont{Z.}~\bibnamefont{Tian}},
  \bibinfo{journal}{Inorganic Chemistry} \textbf{\bibinfo{volume}{59}},
  \bibinfo{pages}{5368} (\bibinfo{year}{2020}).

\bibitem[{\citenamefont{Rau and Gingras}(2019)}]{rau2019frustrated}
\bibinfo{author}{\bibfnamefont{J.~G.} \bibnamefont{Rau}} \bibnamefont{and}
  \bibinfo{author}{\bibfnamefont{M.~J.} \bibnamefont{Gingras}},
  \bibinfo{journal}{Annual Review of Condensed Matter Physics}
  (\bibinfo{year}{2019}).

\bibitem[{\citenamefont{Rau et~al.}(2018)\citenamefont{Rau, Wu, May, Taylor,
  Liu, Higgins, Butch, Ross, Nair, Lumsden et~al.}}]{Rau2018}
\bibinfo{author}{\bibfnamefont{J.~G.} \bibnamefont{Rau}},
  \bibinfo{author}{\bibfnamefont{L.~S.} \bibnamefont{Wu}},
  \bibinfo{author}{\bibfnamefont{A.~F.} \bibnamefont{May}},
  \bibinfo{author}{\bibfnamefont{A.~E.} \bibnamefont{Taylor}},
  \bibinfo{author}{\bibfnamefont{I.-L.} \bibnamefont{Liu}},
  \bibinfo{author}{\bibfnamefont{J.}~\bibnamefont{Higgins}},
  \bibinfo{author}{\bibfnamefont{N.~P.} \bibnamefont{Butch}},
  \bibinfo{author}{\bibfnamefont{K.~A.} \bibnamefont{Ross}},
  \bibinfo{author}{\bibfnamefont{H.~S.} \bibnamefont{Nair}},
  \bibinfo{author}{\bibfnamefont{M.~D.} \bibnamefont{Lumsden}},
  \bibnamefont{et~al.}, \bibinfo{journal}{Journal of Physics: Condensed Matter}
  \textbf{\bibinfo{volume}{30}}, \bibinfo{pages}{1} (\bibinfo{year}{2018}).

\bibitem[{\citenamefont{Wu et~al.}(2019)\citenamefont{Wu, Nikitin, Wang, Zhu,
  Batista, Tsvelik, Samarakoon, Tennant, Brando, Vasylechko et~al.}}]{Wu2019}
\bibinfo{author}{\bibfnamefont{L.~S.} \bibnamefont{Wu}},
  \bibinfo{author}{\bibfnamefont{S.~E.} \bibnamefont{Nikitin}},
  \bibinfo{author}{\bibfnamefont{Z.}~\bibnamefont{Wang}},
  \bibinfo{author}{\bibfnamefont{W.}~\bibnamefont{Zhu}},
  \bibinfo{author}{\bibfnamefont{C.~D.} \bibnamefont{Batista}},
  \bibinfo{author}{\bibfnamefont{A.~M.} \bibnamefont{Tsvelik}},
  \bibinfo{author}{\bibfnamefont{A.~M.} \bibnamefont{Samarakoon}},
  \bibinfo{author}{\bibfnamefont{D.~A.} \bibnamefont{Tennant}},
  \bibinfo{author}{\bibfnamefont{M.}~\bibnamefont{Brando}},
  \bibinfo{author}{\bibfnamefont{L.}~\bibnamefont{Vasylechko}},
  \bibnamefont{et~al.}, \bibinfo{journal}{Nat. Commun.}
  \textbf{\bibinfo{volume}{10}} (\bibinfo{year}{2019}).

\bibitem[{\citenamefont{Hester et~al.}(2019)\citenamefont{Hester, Nair, Reeder,
  Yahne, DeLazzer, Berges, Ziat, Neilson, Aczel, Sala et~al.}}]{Hester2019}
\bibinfo{author}{\bibfnamefont{G.}~\bibnamefont{Hester}},
  \bibinfo{author}{\bibfnamefont{H.~S.} \bibnamefont{Nair}},
  \bibinfo{author}{\bibfnamefont{T.}~\bibnamefont{Reeder}},
  \bibinfo{author}{\bibfnamefont{D.~R.} \bibnamefont{Yahne}},
  \bibinfo{author}{\bibfnamefont{T.~N.} \bibnamefont{DeLazzer}},
  \bibinfo{author}{\bibfnamefont{L.}~\bibnamefont{Berges}},
  \bibinfo{author}{\bibfnamefont{D.}~\bibnamefont{Ziat}},
  \bibinfo{author}{\bibfnamefont{J.~R.} \bibnamefont{Neilson}},
  \bibinfo{author}{\bibfnamefont{A.~A.} \bibnamefont{Aczel}},
  \bibinfo{author}{\bibfnamefont{G.}~\bibnamefont{Sala}}, \bibnamefont{et~al.},
  \bibinfo{journal}{Phys. Rev. Lett.} \textbf{\bibinfo{volume}{123}},
  \bibinfo{pages}{027201} (\bibinfo{year}{2019}).

\bibitem[{\citenamefont{Kitaev}(2006)}]{Kitaev2006}
\bibinfo{author}{\bibfnamefont{A.}~\bibnamefont{Kitaev}},
  \bibinfo{journal}{Annals of Physics} \textbf{\bibinfo{volume}{321}},
  \bibinfo{pages}{2} (\bibinfo{year}{2006}).

\bibitem[{\citenamefont{Jackeli and Khaliullin}(2009)}]{Jackeli2009}
\bibinfo{author}{\bibfnamefont{G.}~\bibnamefont{Jackeli}} \bibnamefont{and}
  \bibinfo{author}{\bibfnamefont{G.}~\bibnamefont{Khaliullin}},
  \bibinfo{journal}{Phys. Rev. Lett.} \textbf{\bibinfo{volume}{102}},
  \bibinfo{pages}{2} (\bibinfo{year}{2009}).

\bibitem[{\citenamefont{Ross et~al.}(2011)\citenamefont{Ross, Savary, Gaulin,
  and Balents}}]{Ross2011}
\bibinfo{author}{\bibfnamefont{K.~A.} \bibnamefont{Ross}},
  \bibinfo{author}{\bibfnamefont{L.}~\bibnamefont{Savary}},
  \bibinfo{author}{\bibfnamefont{B.~D.} \bibnamefont{Gaulin}},
  \bibnamefont{and} \bibinfo{author}{\bibfnamefont{L.}~\bibnamefont{Balents}},
  \bibinfo{journal}{Phys. Rev. X} \textbf{\bibinfo{volume}{1}},
  \bibinfo{pages}{1} (\bibinfo{year}{2011}).

\bibitem[{\citenamefont{Hermele et~al.}(2004)\citenamefont{Hermele, Fisher, and
  Balents}}]{hermele2004pyrochlore}
\bibinfo{author}{\bibfnamefont{M.}~\bibnamefont{Hermele}},
  \bibinfo{author}{\bibfnamefont{M.~P.} \bibnamefont{Fisher}},
  \bibnamefont{and} \bibinfo{author}{\bibfnamefont{L.}~\bibnamefont{Balents}},
  \bibinfo{journal}{Phys. Rev. B.} \textbf{\bibinfo{volume}{69}},
  \bibinfo{pages}{064404} (\bibinfo{year}{2004}).

\bibitem[{\citenamefont{Banerjee et~al.}(2016)\citenamefont{Banerjee, Bridges,
  Yan, Aczel, Li, Stone, Granroth, Lumsden, Yiu, Knolle et~al.}}]{Banerjee2016}
\bibinfo{author}{\bibfnamefont{A.}~\bibnamefont{Banerjee}},
  \bibinfo{author}{\bibfnamefont{C.~A.} \bibnamefont{Bridges}},
  \bibinfo{author}{\bibfnamefont{J.-Q.} \bibnamefont{Yan}},
  \bibinfo{author}{\bibfnamefont{A.~A.} \bibnamefont{Aczel}},
  \bibinfo{author}{\bibfnamefont{L.}~\bibnamefont{Li}},
  \bibinfo{author}{\bibfnamefont{M.~B.} \bibnamefont{Stone}},
  \bibinfo{author}{\bibfnamefont{G.~E.} \bibnamefont{Granroth}},
  \bibinfo{author}{\bibfnamefont{M.~D.} \bibnamefont{Lumsden}},
  \bibinfo{author}{\bibfnamefont{Y.}~\bibnamefont{Yiu}},
  \bibinfo{author}{\bibfnamefont{J.}~\bibnamefont{Knolle}},
  \bibnamefont{et~al.}, \bibinfo{journal}{Nature Materials}
  \textbf{\bibinfo{volume}{15}}, \bibinfo{pages}{733} (\bibinfo{year}{2016}).

\bibitem[{\citenamefont{Choi et~al.}(2019)\citenamefont{Choi, Lee, Lee, Yoon,
  Lee, Park, Ali, Singh, Orain, Kim et~al.}}]{choi2019exotic}
\bibinfo{author}{\bibfnamefont{Y.}~\bibnamefont{Choi}},
  \bibinfo{author}{\bibfnamefont{C.}~\bibnamefont{Lee}},
  \bibinfo{author}{\bibfnamefont{S.}~\bibnamefont{Lee}},
  \bibinfo{author}{\bibfnamefont{S.}~\bibnamefont{Yoon}},
  \bibinfo{author}{\bibfnamefont{W.-J.} \bibnamefont{Lee}},
  \bibinfo{author}{\bibfnamefont{J.}~\bibnamefont{Park}},
  \bibinfo{author}{\bibfnamefont{A.}~\bibnamefont{Ali}},
  \bibinfo{author}{\bibfnamefont{Y.}~\bibnamefont{Singh}},
  \bibinfo{author}{\bibfnamefont{J.-C.} \bibnamefont{Orain}},
  \bibinfo{author}{\bibfnamefont{G.}~\bibnamefont{Kim}}, \bibnamefont{et~al.},
  \bibinfo{journal}{Phys. Rev. Lett.} \textbf{\bibinfo{volume}{122}},
  \bibinfo{pages}{167202} (\bibinfo{year}{2019}).

\bibitem[{\citenamefont{Jackeli and Khaliullin}(2013)}]{Jackeli2013}
\bibinfo{author}{\bibfnamefont{G.}~\bibnamefont{Jackeli}} \bibnamefont{and}
  \bibinfo{author}{\bibfnamefont{G.}~\bibnamefont{Khaliullin}},
  \bibinfo{journal}{Phys. Rev. Lett.} \textbf{\bibinfo{volume}{110}},
  \bibinfo{pages}{097204} (\bibinfo{year}{2013}).

\bibitem[{\citenamefont{{Hwan Chun} et~al.}(2015)\citenamefont{{Hwan Chun},
  Kim, Kim, Zheng, Stoumpos, Malliakas, Mitchell, Mehlawat, Singh, Choi
  et~al.}}]{HwanChun2015}
\bibinfo{author}{\bibfnamefont{S.}~\bibnamefont{{Hwan Chun}}},
  \bibinfo{author}{\bibfnamefont{J.~W.} \bibnamefont{Kim}},
  \bibinfo{author}{\bibfnamefont{J.}~\bibnamefont{Kim}},
  \bibinfo{author}{\bibfnamefont{H.}~\bibnamefont{Zheng}},
  \bibinfo{author}{\bibfnamefont{C.~C.} \bibnamefont{Stoumpos}},
  \bibinfo{author}{\bibfnamefont{C.~D.} \bibnamefont{Malliakas}},
  \bibinfo{author}{\bibfnamefont{J.~F.} \bibnamefont{Mitchell}},
  \bibinfo{author}{\bibfnamefont{K.}~\bibnamefont{Mehlawat}},
  \bibinfo{author}{\bibfnamefont{Y.}~\bibnamefont{Singh}},
  \bibinfo{author}{\bibfnamefont{Y.}~\bibnamefont{Choi}}, \bibnamefont{et~al.},
  \bibinfo{journal}{Nature Physics} \textbf{\bibinfo{volume}{11}},
  \bibinfo{pages}{462} (\bibinfo{year}{2015}).

\bibitem[{\citenamefont{Momma and Izumi}(2011)}]{vesta}
\bibinfo{author}{\bibfnamefont{K.}~\bibnamefont{Momma}} \bibnamefont{and}
  \bibinfo{author}{\bibfnamefont{F.}~\bibnamefont{Izumi}},
  \bibinfo{journal}{Journal of Applied Crystallography}
  \textbf{\bibinfo{volume}{44}}, \bibinfo{pages}{1272} (\bibinfo{year}{2011}).

\bibitem[{\citenamefont{Felsche}(1970)}]{Felsche1970}
\bibinfo{author}{\bibfnamefont{J.}~\bibnamefont{Felsche}},
  \bibinfo{journal}{Journal of The Less-Common Metals}
  \textbf{\bibinfo{volume}{21}}, \bibinfo{pages}{1} (\bibinfo{year}{1970}).

\bibitem[{\citenamefont{Smolin and Shepelev}(1970)}]{smolin1970crystal}
\bibinfo{author}{\bibfnamefont{Y.~I.} \bibnamefont{Smolin}} \bibnamefont{and}
  \bibinfo{author}{\bibfnamefont{Y.~F.} \bibnamefont{Shepelev}},
  \bibinfo{journal}{Acta Crystallographica} \textbf{\bibinfo{volume}{26}},
  \bibinfo{pages}{484} (\bibinfo{year}{1970}).

\bibitem[{\citenamefont{Hester et~al.}(2020)\citenamefont{Hester, DeLazzer,
  Yahne, Sarkis, Zhao, Rivera, Calder, and Ross}}]{Hester2020}
\bibinfo{author}{\bibfnamefont{G.}~\bibnamefont{Hester}},
  \bibinfo{author}{\bibfnamefont{T.~N.} \bibnamefont{DeLazzer}},
  \bibinfo{author}{\bibfnamefont{D.~R.} \bibnamefont{Yahne}},
  \bibinfo{author}{\bibfnamefont{C.~L.} \bibnamefont{Sarkis}},
  \bibinfo{author}{\bibfnamefont{H.~D.} \bibnamefont{Zhao}},
  \bibinfo{author}{\bibfnamefont{J.~A.~R.} \bibnamefont{Rivera}},
  \bibinfo{author}{\bibfnamefont{S.}~\bibnamefont{Calder}}, \bibnamefont{and}
  \bibinfo{author}{\bibfnamefont{K.~A.} \bibnamefont{Ross}}, pp.
  \bibinfo{pages}{1--10} (\bibinfo{year}{2020}), \eprint{arXiv: 2008.00041}.

\bibitem[{\citenamefont{Redhammer and Roth}(2003)}]{redhammer2003beta}
\bibinfo{author}{\bibfnamefont{G.~u.~J.} \bibnamefont{Redhammer}}
  \bibnamefont{and} \bibinfo{author}{\bibfnamefont{G.}~\bibnamefont{Roth}},
  \bibinfo{journal}{Acta Crystallographica Section C: Crystal Structure
  Communications} \textbf{\bibinfo{volume}{59}}, \bibinfo{pages}{i103}
  (\bibinfo{year}{2003}).

\bibitem[{\citenamefont{Kahlenberg and
  Aichholzer}(2014)}]{kahlenberg2014thortveitite}
\bibinfo{author}{\bibfnamefont{V.}~\bibnamefont{Kahlenberg}} \bibnamefont{and}
  \bibinfo{author}{\bibfnamefont{P.}~\bibnamefont{Aichholzer}},
  \bibinfo{journal}{Acta Crystallographica Section E: Structure Reports Online}
  \textbf{\bibinfo{volume}{70}}, \bibinfo{pages}{i34} (\bibinfo{year}{2014}).

\bibitem[{\citenamefont{Ciomaga~Hatnean
  et~al.}(2020)\citenamefont{Ciomaga~Hatnean, Petrenko, Lees, Orton, and
  Balakrishnan}}]{ciomaga2020optical}
\bibinfo{author}{\bibfnamefont{M.}~\bibnamefont{Ciomaga~Hatnean}},
  \bibinfo{author}{\bibfnamefont{O.~A.} \bibnamefont{Petrenko}},
  \bibinfo{author}{\bibfnamefont{M.~R.} \bibnamefont{Lees}},
  \bibinfo{author}{\bibfnamefont{T.~E.} \bibnamefont{Orton}}, \bibnamefont{and}
  \bibinfo{author}{\bibfnamefont{G.}~\bibnamefont{Balakrishnan}},
  \bibinfo{journal}{Crystal Growth \& Design} \textbf{\bibinfo{volume}{20}},
  \bibinfo{pages}{6636} (\bibinfo{year}{2020}).

\bibitem[{\citenamefont{Maqsood et~al.}(1979)\citenamefont{Maqsood, Wanklyn,
  and Garton}}]{Maqsood1979}
\bibinfo{author}{\bibfnamefont{A.}~\bibnamefont{Maqsood}},
  \bibinfo{author}{\bibfnamefont{B.~M.} \bibnamefont{Wanklyn}},
  \bibnamefont{and} \bibinfo{author}{\bibfnamefont{G.}~\bibnamefont{Garton}},
  \bibinfo{journal}{Journal of Crystal Growth} \textbf{\bibinfo{volume}{46}},
  \bibinfo{pages}{671} (\bibinfo{year}{1979}).

\bibitem[{\citenamefont{Maqsood}(1981)}]{Maqsood1981}
\bibinfo{author}{\bibfnamefont{A.}~\bibnamefont{Maqsood}},
  \bibinfo{journal}{Journal of Materials Science}
  \textbf{\bibinfo{volume}{16}}, \bibinfo{pages}{2198} (\bibinfo{year}{1981}).

\bibitem[{\citenamefont{Ameer et~al.}(2012)\citenamefont{Ameer, Faraz, Maqsood,
  and Ahmad}}]{Ameer2012}
\bibinfo{author}{\bibfnamefont{S.}~\bibnamefont{Ameer}},
  \bibinfo{author}{\bibfnamefont{A.}~\bibnamefont{Faraz}},
  \bibinfo{author}{\bibfnamefont{A.}~\bibnamefont{Maqsood}}, \bibnamefont{and}
  \bibinfo{author}{\bibfnamefont{N.~M.} \bibnamefont{Ahmad}},
  \bibinfo{journal}{Journal of Nano Research} \textbf{\bibinfo{volume}{17}},
  \bibinfo{pages}{85} (\bibinfo{year}{2012}).

\bibitem[{\citenamefont{Rodr{\'\i}guez-Carvajal}(1993)}]{fullprof}
\bibinfo{author}{\bibfnamefont{J.}~\bibnamefont{Rodr{\'\i}guez-Carvajal}},
  \bibinfo{journal}{Physica B: Condensed Matter}
  \textbf{\bibinfo{volume}{192}}, \bibinfo{pages}{55} (\bibinfo{year}{1993}).

\bibitem[{\citenamefont{Wills}(2009)}]{sarah1}
\bibinfo{author}{\bibfnamefont{A.}~\bibnamefont{Wills}}, \bibinfo{journal}{Z
  Kristallogr} \textbf{\bibinfo{volume}{30}}, \bibinfo{pages}{39}
  (\bibinfo{year}{2009}).

\bibitem[{\citenamefont{Wills}(2000)}]{sarah2}
\bibinfo{author}{\bibfnamefont{A.}~\bibnamefont{Wills}},
  \bibinfo{journal}{Physica B: Condensed Matter}
  \textbf{\bibinfo{volume}{276}}, \bibinfo{pages}{680} (\bibinfo{year}{2000}).

\bibitem[{\citenamefont{Fouet et~al.}(2001)\citenamefont{Fouet, Sindzingre, and
  Lhuillier}}]{Fouet2001}
\bibinfo{author}{\bibfnamefont{J.~B.} \bibnamefont{Fouet}},
  \bibinfo{author}{\bibfnamefont{P.}~\bibnamefont{Sindzingre}},
  \bibnamefont{and}
  \bibinfo{author}{\bibfnamefont{C.}~\bibnamefont{Lhuillier}},
  \bibinfo{journal}{European Physical Journal B} \textbf{\bibinfo{volume}{20}},
  \bibinfo{pages}{241} (\bibinfo{year}{2001}).

\bibitem[{\citenamefont{Rau and Gingras}(2018)}]{Rau2018Yb}
\bibinfo{author}{\bibfnamefont{J.~G.} \bibnamefont{Rau}} \bibnamefont{and}
  \bibinfo{author}{\bibfnamefont{M.~J.~P.} \bibnamefont{Gingras}},
  \bibinfo{journal}{Phys. Rev. B} \textbf{\bibinfo{volume}{98}},
  \bibinfo{pages}{054408} (\bibinfo{year}{2018}).

\bibitem[{\citenamefont{Nair et~al.}(2019)\citenamefont{Nair, DeLazzer, Reeder,
  Sikorski, Hester, and Ross}}]{Nair2019}
\bibinfo{author}{\bibfnamefont{H.}~\bibnamefont{Nair}},
  \bibinfo{author}{\bibfnamefont{T.}~\bibnamefont{DeLazzer}},
  \bibinfo{author}{\bibfnamefont{T.}~\bibnamefont{Reeder}},
  \bibinfo{author}{\bibfnamefont{A.}~\bibnamefont{Sikorski}},
  \bibinfo{author}{\bibfnamefont{G.}~\bibnamefont{Hester}}, \bibnamefont{and}
  \bibinfo{author}{\bibfnamefont{K.}~\bibnamefont{Ross}},
  \bibinfo{journal}{Crystals} \textbf{\bibinfo{volume}{9}}, \bibinfo{pages}{10}
  (\bibinfo{year}{2019}).

\bibitem[{\citenamefont{Sarkis et~al.}(2020)\citenamefont{Sarkis, Rau,
  Sanjeewa, Powell, Kolis, Marbey, Hill, Rodriguez-Rivera, Nair, Yahne
  et~al.}}]{sarkis2020}
\bibinfo{author}{\bibfnamefont{C.}~\bibnamefont{Sarkis}},
  \bibinfo{author}{\bibfnamefont{J.~G.} \bibnamefont{Rau}},
  \bibinfo{author}{\bibfnamefont{L.}~\bibnamefont{Sanjeewa}},
  \bibinfo{author}{\bibfnamefont{M.}~\bibnamefont{Powell}},
  \bibinfo{author}{\bibfnamefont{J.}~\bibnamefont{Kolis}},
  \bibinfo{author}{\bibfnamefont{J.}~\bibnamefont{Marbey}},
  \bibinfo{author}{\bibfnamefont{S.}~\bibnamefont{Hill}},
  \bibinfo{author}{\bibfnamefont{J.}~\bibnamefont{Rodriguez-Rivera}},
  \bibinfo{author}{\bibfnamefont{H.}~\bibnamefont{Nair}},
  \bibinfo{author}{\bibfnamefont{D.}~\bibnamefont{Yahne}},
  \bibnamefont{et~al.}, \bibinfo{journal}{Physical Review B}
  \textbf{\bibinfo{volume}{102}}, \bibinfo{pages}{134418}
  (\bibinfo{year}{2020}).

\end{thebibliography}

\end{document}